**Performance of multilabel machine learning models and risk stratification schemas for predicting stroke and bleeding risk in patients with non-valvular atrial fibrillation**

Authors: Juan Lu, MPE[1,2,3], Rebecca Hutchens, MBBS[2,4], Joseph Hung, MBBS[2], Mohammed Bennamoun, PhD[1], Brendan McQuillan, PhD[2,4], Tom Briffa, PhD[5], Ferdous Sohel, PhD[6], Kevin Murray, PhD[5], Jonathon Stewart, MBBS[2,3], Benjamin Chow, MD[7], Frank Sanfilippo, PhD[5], Girish Dwivedi, PhD[2,3,8]

[1] Department of Computer Science and Software Engineering, The University of Western Australia, Australia.

[2] Medical School, The University of Western Australia, Perth, Australia

[3] Harry Perkins Institute of Medical Research, Perth, Australia

[4] Sir Charles Gairdner Hospital, Perth, WA, Australia

[5] School of Population and Global Health, The University of Western Australia, Australia.

[6] Discipline of Information Technology, Murdoch University, Perth, Australia.

[7] Department of Medicine, Division of Cardiology, University of Ottawa Heart Institute, Ottawa, Canada

[8] Cardiology Department, Fiona Stanley Hospital, Perth, Australia.

Short Title: Machine learning models to predict outcomes in AF

**Corresponding author:**

Dr. Girish Dwivedi, MD, FRCPC, FACC, FASNC, FSCCT

Wesfarmers Chair in Cardiology, Harry Perkins Institute of Medical Research, The University of Western Australia

Consultant Cardiologist at Fiona Stanley Hospital, Murdoch, 6009, WA, Australia

Adjunct Professor of Cardiology, University of Ottawa, Canada

Adjunct Research Professor-School of Biomedical Sciences at Curtin University, WA, Australia

Email:girish.dwivedi@perkins.uwa.edu.au

T: +61 8 6488 8181

**Abstract**

**Objective**

Appropriate antithrombotic therapy for patients with atrial fibrillation (AF) requires assessment of ischemic stroke and bleeding risks. However, risk stratification schemas such as $CHA_2DS_2$-VASc and HAS-BLED have modest predictive capacity for patients with AF. Machine learning (ML) techniques may improve predictive performance and support decision-making for appropriate antithrombotic therapy. We compared the performance of multilabel ML models with the currently used risk scores for predicting outcomes in AF patients.

**Materials and Methods**

This was a retrospective cohort study of 9670 patients, mean age 76.9 years, 46% women, who were hospitalized with non-valvular AF, and had 1-year follow-up. The primary outcome was ischemic stroke and major bleeding admission. The secondary outcomes were all-cause death and event-free survival. Discriminant power of ML models were compared with clinical risk scores by area under the curve (AUC). Risk stratification was assessed using net reclassification index.

**Results**

Multilabel gradient boosting machine provided the best discriminant power for stroke, major bleeding and death (AUC = 0.685, 0.709, and 0.765 respectively) compared to other ML models. It provided modest performance improvement for stroke compared to $CHA_2DS_2$-VASc (AUC = 0.652), but significantly improved major bleeding prediction compared to HAS-BLED (AUC = 0.522). It also had much greater discriminant power for death compared with $CHA_2DS_2$-VASc (AUC = 0.606). Also, models identified additional risk features (such as hemoglobin level, renal function etc.) for each outcome.

**Conclusions**

Multilabel ML models can outperform clinical risk stratification scores for predicting the risk of major bleeding and death in non-valvular AF patients.

# INTRODUCTION

Atrial fibrillation (AF) is a well-recognized risk factor for ischemic stroke and is associated with a substantial risk of morbidity and mortality.[1] Oral anticoagulant (OAC) therapy is widely used to reduce the risk of ischemic stroke but at the expense of increasing patients' risk of bleeding.[2] Different risk stratification schemas have been developed, which serve as decision-making tools for clinicians to help administer antithrombotic therapy to patients with AF while balancing the stroke and bleeding risks.[3] The widely used $CHA_2DS_2$-VASc score[4] estimates a patients' annual stroke risk based on identified risk factors. ATRIA,[5] ORBIT[6] and HAS-BLED[7] scores are widely used for bleeding risk assessment.

However, these stratification schemas only allow for cumulative assessments of risk factors, which may oversimplify their relationships. Moreover, there are a number of risk factors in common between the bleeding risk scores and the $CHA_2DS_2$-VASc score.[8] This means patients with a high risk of stroke are also at increased risk of bleeding, which makes it hard to derive anticoagulation decisions.

Machine learning (ML) has broadly been applied in patients' cardiovascular risk prediction, the discovery of knowledge and optimal decision making using clinical and administrative data. In many but not all cases, ML provided better results than the traditional methods.[9–12] ML may improve patients' risk of stroke and bleeding prediction as it allows for nonlinear associations and is better performed at extracting additional information from continuous variables (such as age). Additionally, multilabel ML can predict different patients' outcomes at the same time. This may help clinicians in balancing patients' risks while making the decisions on anticoagulant therapy. Lastly, in addition to potential performance improvement, ML can rank features from the input data, which may suggest other important risk factors.[13]

In this study, we aimed to investigate whether ML can outperform the traditional stratification schemas for ischemic stroke and major bleeding prediction in patients with AF and pick up important risk factors other than the ones in $CHA_2DS_2$-VASc and bleeding risk scores. Figure 1 depicts the proposed framework. The developed model could be used to assist clinicians in prescribing antithrombotic therapy for patients with AF.

## METHODS

**Study Population**

This was a retrospective cohort study in a secondary care setting of all adults with a history of AF or atrial flutter admitted to Sir Charles Gairdner and Osborne Park teaching hospitals in Perth, Western Australia, between Jan 1, 2009 and the March 31, 2016. The selection and details of the study cohort have been previously described.[14] Patients with AF listed as a primary, secondary or past medical history were included in the study cohort after excluding patients with mitral valve stenosis and/or prosthetic heart valve history. Patients' index (first) admission during the study period was used for data analysis. This study was approved by the Institutional Review Board of the Western Australian Department of Health Western Australia (201610.04) and Sir Charles Gairdner Hospital (2011-154).

**Data sources**

From CGMS, we extracted demographic information, past medical history and discharge medications (see Table 1). Laboratory results recorded during the admission were extracted from PathWest database, including full blood picture, urea and electrolytes and liver function test results. We calculated patients' estimated glomerular filtration rate (eGFR) results using the Chronic Kidney Disease Epidemiology Collaboration (CKD-EPI) equation.[15]

**Tables**

**Table 1. Cohort Characteristics**

| **Patient characteristics**, n (%) unless otherwise specified | Total |
|---|---|
| Number of patients | 9670 |
| **Demographics** | |
| Age, mean (SD), years | 76.9 (12.13) |
| Age group, years | |
| < 65 | 1470 (15.2) |
| 65 - 74 | 1977 (20.4) |
| 75 + | 6223 (64.4) |
| Male | 5234 (54.1) |
| Rural | 1167 (12.1) |
| $CHA_2DS_2$-VASc score, mean (SD) | 3.82 (1.81) |
| $CHA_2DS_2$-VASc categories n (%) | |
| 0 | 356 (3.7) |
| 1 | 706 (7.3) |
| ≥2 | 8608 (89.0) |
| HAS-BLED score, mean (SD) | 2.29 (1.04) |
| HAS-BLED categories n (%) | |
| 0 | 420 (4.3) |
| 1 | 1688 (17.5) |
| ≥2 | 7562 (78.2) |
| ATRIA score, mean (SD) | 2.23 (1.55) |
| 0 - 3 | 8531 (88.2) |
| 4 | 468 (4.8) |
| ≥5 | 671 (6.9) |
| ORBIT score, mean (SD) | 1.64 (1.18) |
| 0 - 2 | 7949 (82.2) |
| 3 | 995 (10.3) |
| ≥4 | 726 (7.5) |
| Cardiology admission | 1782 (18.4) |
| **Medical history** | |
| Hypertension | 5423 (56.1) |
| Heart failure | 2471 (25.6) |
| Diabetes mellitus | 2280 (23.6) |
| Previous thromboembolism[f] | 1981 (20.5) |
| Vascular disease | 3857 (39.9) |
| Bleeding history | 1650 (17.1) |
| Cognitive impairment[a] | 1208 (12.5) |
| Dementia | 726 (7.5) |
| Fall risk | 1528 (15.8) |
| Depression | 917 (9.5) |
| Liver Impairment | 636 (6.6) |
| Arthritis | 2775 (28.7) |
| Deep vein thrombus / Pulmonary embolism | 567 (5.9) |
| Anaemia[b] | 5935 (61.4) |
| Number of comorbidities, mean (SD) | 2.90 (1.74) |
| **Medication treatment** | |
| Oral anticoagulant [c] | 3404 (35.2) |
| Vitamin K antagonists | 2869 (29.7) |
| Non-vitamin K antagonist Oral anticoagulant N | 535 (5.5) |
| Antiplatelet drug | 4404 (45.5) |
| Rate control drug | 6831 (70.6) |
| Rhythm control drug | 1379 (14.3) |
| Proton pump inhibitors | 4582 (47.4) |
| Polypharmacy[d] | 1281 (13.3) |
| **Test result** | |
| Hemoglobin worst result mean g/L (SD) | 108.4 (33.8) |
| CKD stages based on eGFR[e] | |
| 1 (>= 90) | 1213 (12.6) |
| 2 (60 - 90) | 3783 (39.1) |
| 3 (30 - 60) | 3327 (34.4) |
| 4 (15 - 30) | 962 (10.0) |
| 5 (< 15) | 385 (4.0) |

Abbreviations: SD, standard deviation; TIA, transient ischemic attack; DVT, deep vein thrombosis; PE, pulmonary embolus; eGFR, estimated glomerular filtration rate; NOAC, non-vitamin K oral anticoagulant; VKA, vitamin K antagonists; CGMS, Clinical Governance Management System; CKD, chronic kidney disease; eGFR, estimated Glomerular Filtration Rate.

[a] Cognitive impairment was defined as delirium, dementia and cognitive impairment cause not specified (e.g. confusion, short-term memory impairment, and cognitive impairment)

[b] Anaemia was defined by using the worst hemoglobin result and WHO definition of anaemia

[c] OAC was defined as patients discharged on a vitamin K antagonist (VKA) or a non-vitamin K dependent oral anticoagulant (NOAC)

[d] Polypharmacy was defined as use of five or more medications at discharge

[e] Missing values in eGFR were treated as normal and put into the group "1" Death was defined as all-cause death except for ischemic stroke and major bleeding.

[f] Previous thromboembolism defined as stroke/trans ischaemic attack/arterial embolism

**Predictors and outcomes**

The risk variables used to calculate $CHA_2DS_2$-VASc and bleeding risk scores (HAS-BLED, ATRIA, ORBIT) for each patient were obtained using electronic discharge summaries in CGMS and laboratory results (see Supplemental table I-IV). For HAS-BLED, unstable international normalized ratios were omitted as these could not be reliably ascertained.

Index AF cases were linked to the routinely collected and audited WA Hospital Morbidity Data Collection (HMDC) and the WA Death Registry to assess for specified outcomes following their admission.[16] We allowed for one-year follow-up after hospital discharge for all patients using the linked HMDC with the last available date of follow-up being 30/06/16. The primary outcomes of interest identified using International Classification of Diseases - Tenth Revision - Australian Modification (ICD-10-AM) codes were an ischemic stroke (as principal diagnosis) and major bleeding defined as admission for gastrointestinal hemorrhage, intracranial hemorrhage (hemorrhagic stroke) or anticoagulant-related bleeding as a principal or secondary discharge diagnosis. (Supplemental table V) Secondary outcomes were all-cause death and event-free survival comprising patients who were event-free and still alive at 1-year follow-up.

We analyzed $CHA_2DS_2$-VASc, HAS-BLED, ATRIA and ORBIT as continuous risk scores using logistic regression models. We compiled the ML models on clinical, medication and laboratory variables available in the cohort dataset (see Table 1). These variables included the individual risk factors used in the calculation of stroke and bleeding risk scores and were included as binary variables (yes/no) excepting age which was treated as a continuous variable. Other variables entered in the ML models included year of cohort entry, rural residence, specialty admission (Cardiology), length of hospital stay, cognitive impairment/dementia, falls risk, depression, use of antiplatelet agents, OACs (warfarin or non-vitamin-K dependent OAC [NOAC])), rate and/or rhythm control drugs, gastro-protective agents, polypharmacy (5 or more medications), anaemia (WHO criteria), lowest hemoglobin result (g/L), and chronic kidney disease (CKD) stage (1-5) based on eGFR (see Table 1 and Supplemental Table VI). We standardized continuous variables by removing the mean and scaling to unit variance to avoid potential bias.

**Missing data**

All the variables included in the ML model were complete, except for eGFR (4.7% missing). Missing eGFR values were imputed to be ⩾ 90, as kidney function tests including eGFR are performed in all admitted patients except when they are known to be normal.

**ML Algorithms for multilabel classification**

We aimed to predict patients' risk of ischemic stroke, major bleeding, or death. To address this multilabel problem, we used classifier chain[17] together with traditional classification algorithms: support vector machine (SVM), gradient boosting machine (GBM), and multi-layer neural networks (MLNN), which have been identified to be powerful classification techniques.[18–20] These binary classification algorithms are representative algorithms of different types of machine learning. SVM is an instance-based algorithm, GBM is an ensemble algorithm, and MLNN is a neural network algorithm.[21] Figure 3 showed the proposed multilabel pipeline for SVM and GBM. We constructed a chain of binary classifiers, one for each label (outcome). The $k^{th}$ classifier uses the predictions of previous classifiers. Different from methods such as binary relevance and OneVsRest, classifier chain takes the label correlation into account. We experimented different ordering of the classes (outcomes) as well as an ensemble of 10 randomly ordered classifier chains by averaging the binary predictions of the chains. The optimal ordering of the classes is presented in Figure 3. MLNN supports multilabel classification itself, we thus did not wrap it in the classifiers chain. The output includes four labels based on patients' 1-year outcomes of ischemic stroke, major bleeding, all-cause death or event-free survival.

SVM is a popular binary classification technique. It builds a hyperplane between two classes, with the maximal distance between the hyperplane and the nearest point. GBM[22] creates a sequence of weak learners (i.e. decision trees) to minimize the loss function. Initially, the first tree was fitted to the data, and the best split points were chosen to minimize the loss, then trees were added one at a time using gradient descent procedure to continuously minimize the loss. MLNN is another widely applied classification technique. The MLNN model consists of a series of simple interconnected neurons representing a nonlinear mapping between the input and the output vectors.[23] We split the data set into training and testing sets, with a ratio of 0.75: 0.25. The training set was further divided into training and validation sets, in the ratio of 9:1. The detailed parameter setting for each model is described in supplementary material.

**Variables Relative Importance**

As it is not practical for clinicians to collect more than 40 features of a patient at one time, to ensure clinical relevance, we calculated the impurity-based feature importance of selected features for the prediction of ischemic stroke and major bleeding as distinct binary classification outcomes from GBM. For each decision tree (weak learner), it thus measured each variable's cumulative influence in splitting the tree and minimizing the entropy (loss) function. The influence of each variable was further averaged over all of the trees, and was the feature importance score for each variable.[24] Features used at the top of the tree contributed a larger fraction of the input samples to the final predictive model, and the expected fraction they contributed was used as an estimate of the relative importance of the features.

**Performance metrics**

The AUC or C-index with 95% confidence intervals (CIs) from the regression models were used to assess the discriminant power of logistic and ML models for the prediction of ischemic stroke, major bleeding, all-cause death, and event-free survival. The calibration was assessed with the Brier score.[25] To ensure the clinical utility, an ideal model is expected to be highly sensitive to identify patients' risks. The thresholds were selected accordingly to 99% sensitivity for stroke, bleeding and death based on AUC, and a slightly lower sensitivity was also allowed if there were no matching thresholds. We also reported F1-score, precision, and recall metrics using the selected thresholds. As a sensitivity analysis, we also performed the same analyses on patients who were not on OAC therapy. In addition, we assessed the net reclassification index (NRI)[26] for ischemic stroke and major bleeding obtained by the best-performed ML model compared to the standard clinical risk scores. For this purpose, we estimated a patient's annual stroke risk based on their $CHA_2DS_2$-VASc score as being low ($< 1.0\%$), moderate ($\geqslant 1.0\%$ to $< 2.0\%$), or high ($\geq 2.0\%$).[27] A similar approach was applied for annual major bleeding risk with patients collapsed into two risk categories being low ($<4\%$) or high ($\geq 4\%$) based on a threshold HAS-BLED score $\geq 3$[7]. We calculated the NRI of best-performed ML models and compared them with the $CHA_2DS_2$-VASc score and HAS-BLED risk schemas.

The performance of the ML prediction algorithms was assessed by calculating the performance metrics for each outcome. We randomly split the dataset into training and testing sets and calculated the performance scores and 95% CI from training and evaluating the models 50 times. We considered the difference as statistically significant if there is no overlap between 95% CIs.

**Software Packages**

All analyses and model building work were performed in Python, v3.6 and SAS v9.4. We used the Keras Library v2.2.4 with Tensor Flow back end v1.12.0 to implement the MLNN model. The SVM and GBM models were implemented using Scikit-learn library v0.20.2.

**RESULTS**

Of 11,294 patients with non-valvular AF, 9670 (85.6%) satisfied the study inclusion criteria and had one-year follow-up available after index AF admission, and survived to 7 days post-discharge (blanking period) without a stroke or major bleeding event. (see Figure 2 for study schema). Overall, 167 (1.7%) patients had an ischemic stroke event, 430 (4.4%) had a major bleeding event, and 1912 (19.8%) patients died during 1-year follow-up. The mean age (SD) of the study cohort was 76.9 (12.1) years, and 54.0% were men. Table 1 shows the baseline clinical characteristics of the cohort and their discharge medications, with 35.2% prescribed OAC therapy at discharge. Their mean (SD) $CHA_2DS_2$-VASc score was 3.82 (1.81), the mean (SD) HAS-BLED score was 2.29 (1.04).

Table 2 shows the AUC (95% CIs) of the $CHA_2DS_2$-VASc, HAS-BLED, ATIRA and ORBIT scores for ischemic stroke or major bleeding outcomes and compares it to the performance of multilabel ML models, which used all available predictor variables (shown in Table 1). In the total AF cohort, GBM had the highest discriminant power (AUC) for both outcomes. However, GBM was only marginally better than the $CHA_2DS_2$-VASc score for predicting ischemic stroke (0.685 vs 0.652 respectively), while largely outperforming the best of the bleeding risk scores (ATRIA) for classifying major bleeding risk (0.709 vs 0.562, respectively). Figure 4A shows the AUC of GBM and $CHA_2DS_2$-VASc and HAS-BLED scores predicting ischemic stroke and major bleeding. Among multilabel ML models, GBM classifier chain had the highest AUCs for all-cause death and survival outcomes (0.765 and 0.738, respectively) and largely outperformed the clinical risk scores (see Table 2). For ischemic stroke, the NRI of GBM was small compared with the $CHA_2DS_2$-VASc score. (NRI = 3.2%; 95% CI 0.9% to 5.6%) while GBM showed a large NRI for major bleeding. (NRI = 22.8%; 95% 21.3% to 24.3%).

The second section of Table 2 shows the AUC results of the sensitivity analysis considering only patients (n= 6266) not on OAC therapy. For the prediction of ischemic stroke, none of the ML models performed significantly better than the $CHA_2DS_2$-VASc score. The lower 95% CI of the AUC for multilabel ML models overlap the upper 95% CI of the $CHA_2DS_2$-VASc AUC. For prediction of major bleeding, the AUC of the GBM model was lower than the value in the total cohort (0.631 versus 0.709 respectively) but still significantly better than the performance of the ATRIA score (0.539). Figure 4B compares the AUC of GBM and $CHA_2DS_2$-VASc and HAS-BLED scores predicting ischemic stroke and major bleeding for patients not on OAC therapy. In this sub-cohort, ML models still significantly outperformed the clinical risk scores for predicting death and survival outcomes with all AUC ≥0.7. After excluding patients on OAC therapy, the NRI of GBM compared to $CHA_2DS_2$-VASc was small (NRI=3.2%, 95% CI 1% to 5.4%) for predicting ischemic stroke risk, but there was still a modest improvement for prediction of major bleeding risk (NRI = 11.6%; 95% CI 9.2% to 14.1%).

**Table 2.** Area under the receiver operating receiver curve (AUC) and 95% confidence interval (CI) for multilabel ML models compared to CHA$_2$DS$_2$-VASc, HAS-BLED, ATRIA and ORBIT in the overall AF cohort (n=9670) and patients not on oral anticoagulant therapy (n= 6266)

| Outcome | CHA$_2$DS$_2$-VASc[a] | HAS-BLED | ATRIA | ORBIT | SVM | MLNN | GBM |
|---|---|---|---|---|---|---|---|
| **Overall AF cohort (n = 9670)** | | | | | | | |
| Ischemic stroke | 0.652 (0.642, 0.662) | NA | NA | NA | 0.668 (0.659, 0.678) | 0.661 (0.650, 0.672) | 0.685 (0.676, 0.694) |
| Major bleeding | NA | 0.522 (0.516, 0.529) | 0.562 (0.554, 0.570) | 0.511 (0.502, 0.521) | 0.666 (0.661, 0.670) | 0.664 (0.654, 0.674) | 0.709 (0.703, 0.716) |
| All-cause death | 0.606 (0.603, 0.610) | 0.557 (0.555, 0.560) | 0.609 (0.607, 0.612) | 0.611 (0.607, 0.614) | 0.750 (0.747, 0.752) | 0.742 (0.738, 0.756) | 0.765 (0.763, 0.768) |
| Event-free survival | 0.610 (0.606, 0.613) | 0.559 (0.557, 0.562) | 0.607 (0.604, 0.610) | 0.601 (0.597, 0.604) | 0.728 (0.725, 0.731) | 0.726 (0.722, 0.730) | 0.738 (0.735, 0.741) |
| Macro average | NA | NA | NA | NA | 0.685 (0.683, 0.691) | 0.680 (0.674, 0.686) | 0.724 (0.721, 0.727) |
| **Patients not on oral anticoagulant therapy (n = 6266)** | | | | | | | |
| Ischemic stroke | 0.651 (0.638, 0.663) | NA | NA | NA | 0.643 (0.631, 0.654) | 0.634 (0.620, 0.648) | 0.667 (0.660, 0.675) |
| Major bleeding | NA | 0.515 (0.505, 0.525) | 0.539 (0.526, 0.552) | 0.513 (0.501, 0.525) | 0.621 (0.612, 0.632) | 0.600 (0.588, 0.613) | 0.631 (0.620, 0.641) |
| All-cause death | 0.598 (0.593, 0.602) | 0.523 (0.518, 0.528) | 0.589 (0.586, 0.593) | 0.570 (0.565, 0.574) | 0.737 (0.734, 0.740) | 0.725 (0.720, 0.729) | 0.739 (0.736, 0.742) |
| Event-free survival | 0.599 (0.595, 0.602) | 0.526 (0.522, 0.529) | 0.588 (0.585, 0.591) | 0.567 (0.563, 0.570) | 0.724 (0.721, 0.727) | 0.703 (0.698, 0.708) | 0.727 (0.724, 0.730) |
| Macro average | NA | NA | NA | NA | 0.644 (0.640, 0.648) | 0.660 (0.653, 0.666) | 0.69 (0.687, 0.696) |

Abbreviations: NA, not applicable

[a] Average AUC achieved by logistic regression models with CHA$_2$DS$_2$-VASc, HAS-BLED, ATIRA, and ORBIT scores as continuous variables.

All the results were averaged results calculated from running the model on 50 random split data samples.

Table 3 compares the Brier score, precision, recall and F1-score of risk stratification scores to the ML models on the entire AF cohort as well as the sub cohort on patients who are not on oral anticoagulant therapy. The detailed confidence interval for each model is presented in the supplementary. Due to the predefined 99% sensitivity, the precision and F1-score of ischemic stroke, major bleeding and all-cause death are low as expected. The class imbalance of ischemic stroke and major bleeding has further lowered the precision and F1-score. Multilabel machine learning models achieved comparable or greater precision and F1 score comparing to clinical scores focusing on single risk.

**Table 3.** Precision, recall (sensitivity) and F1 score of ML models compared to $CHA_2DS_2$-VASc, HAS-BLED, ATRIA and ORBIT in the overall AF cohort (n=9670) and patients not on oral anticoagulant therapy (n= 6266).

| Outcome | $CHA_2DS_2$-VASc[a] | HAS-BLED | ATRIA | ORBIT | SVM | MLNN | GBM |
|---|---|---|---|---|---|---|---|
| **Ischemic stroke in the total cohort** | | | | | | | |
| Precision | 0.020 | NA | NA | NA | 0.020 | 0.020 | 0.020 |
| Recall* | 0.936 | NA | NA | NA | 0.976 | 0.976 | 0.976 |
| F1-score | 0.040 | NA | NA | NA | 0.040 | 0.039 | 0.040 |
| Brier score | 0.233 | NA | NA | NA | 0.017 | 0.017 | 0.017 |
| **Major bleeding in the total cohort** | | | | | | | |
| Precision | NA | 0.047 | 0.049 | 0.046 | 0.047 | 0.047 | 0.048 |
| Recall* | NA | 0.942 | 0.915 | 0.882 | 0.983 | 0.983 | 0.983 |
| F1-score | NA | 0.089 | 0.091 | 0.088 | 0.090 | 0.090 | 0.091 |
| Brier score | NA | 0.249 | 0.247 | 0.249 | 0.043 | 0.043 | 0.043 |
| **All-cause death in the total cohort** | | | | | | | |
| Precision | 0.211 | 0.204 | 0.219 | 0.215 | 0.223 | 0.224 | 0.225 |
| Recall* | 0.961 | 0.965 | 0.92 | 0.928 | 0.989 | 0.989 | 0.989 |
| F1-score | 0.345 | 0.337 | 0.353 | 0.349 | 0.354 | 0.365 | 0.367 |
| Brier score | 0.241 | 0.247 | 0.240 | 0.241 | 0.143 | 0.143 | 0.155 |
| **Event-free survival in the total cohort** | | | | | | | |
| Precision | 0.812 | 0.791 | 0.812 | 0.822 | 0.867 | 0.877 | 0.852 |
| Recall* | 0.631 | 0.592 | 0.571 | 0.543 | 0.672 | 0.594 | 0.720 |
| F1-score | 0.704 | 0.677 | 0.670 | 0.654 | 0.753 | 0.712 | 0.781 |
| Brier score | 0.240 | 0.247 | 0.241 | 0.242 | 0.166 | 0.165 | 0.178 |
| **Macro average in the total cohort** | | | | | | | |
| Precision | NA | NA | NA | NA | 0.286 | 0.292 | 0.287 |
| Recall* | NA | NA | NA | NA | 0.890 | 0.883 | 0.917 |
| F1-score | NA | NA | NA | NA | 0.296 | 0.299 | 0.320 |
| **Ischemic stroke in the patients who are not on oral anticoagulants55** | | | | | | | |
| Precision | 0.024 | NA | NA | NA | 0.022 | 0.022 | 0.022 |
| Recall* | 0.936 | NA | NA | NA | 0.967 | 0.978 | 0.967 |
| F1-score | 0.047 | NA | NA | NA | 0.043 | 0.042 | 0.043 |
| Brier score | 0.232 | NA | NA | NA | 0.019 | 0.020 | 0.019 |
| **Major bleeding in the patients who are not on oral anticoagulants** | | | | | | | |
| Precision | NA | 0.029 | 0.029 | 0.028 | 0.029 | 0.029 | 0.03 |
| Recall* | NA | 0.925 | 0.884 | 0.899 | 0.977 | 0.977 | 0.977 |
| F1-score | NA | 0.055 | 0.057 | 0.054 | 0.056 | 0.056 | 0.057 |
| Brier score | NA | 0.250 | 0.248 | 0.249 | 0.028 | 0.028 | 0.028 |
| **All-cause death in the patients who are not on oral anticoagulants** | | | | | | | |
| Precision | 0.255 | 0.249 | 0.264 | 0.252 | 0.261 | 0.269 | 0.271 |
| Recall* | 0.972 | 0.948 | 0.919 | 0.936 | 0.989 | 0.985 | 0.989 |
| F1-score | 0.404 | 0.396 | 0.410 | 0.396 | 0.412 | 0.422 | 0.426 |
| Brier score | 0.242 | 0.250 | 0.244 | 0.247 | 0.170 | 0.168 | 0.181 |
| **Event free survival in the patients who are not on oral anticoagulants** | | | | | | | |
| Precision | 0.788 | 0.739 | 0.773 | 0.78 | 0.831 | 0.843 | 0.822 |
| Recall* | 0.485 | 0.514 | 0.534 | 0.422 | 0.58 | 0.589 | 0.712 |
| F1-score | 0.595 | 0.606 | 0.632 | 0.548 | 0.683 | 0.693 | 0.763 |
| Brier score | 0.242 | 0.249 | 0.244 | 0.247 | 0.184 | 0.181 | 0.197 |
| **Macro average in the patients who are not on oral anticoagulants** | | | | | | | |
| Precision | NA | NA | NA | NA | 0.285 | 0.291 | 0.286 |
| Recall* | NA | NA | NA | NA | 0.878 | 0.880 | 0.911 |
| F1-score | NA | NA | NA | NA | 0.298 | 0.303 | 0.322 |

Abbreviations: NA, not applicable

[a] Average precision, recall and F1 score achieved by logistic regression models with $CHA_2DS_2$-VASc, HAS-BLED, ATRIA, ORBIT scores as continuous variables.

* Recall is also known as sensitivity.

All the results were averaged results calculated from running the model on 50 random split data samples.

**Ranking risk predictors**

Figure 5 shows the top 25 most important risk variables shown in descending rank order for prediction of ischemic stroke (Figure 5A), major bleeding (Figure 5C), and all-cause death (Figure 5E) identified by the GBM model in the total cohort. The same risk variables ranked in the order of importance for the sub-cohort not on OAC therapy are shown for comparison (Figure 5B, 5D, and 5F). For ischemic stroke (Figure 5A), the most important risk features (in ranked order) were prior stroke/TIA/embolism history, age (as a continuous variable), hemoglobin level, length of hospital stay, comorbidity number, eGFR grade and prior bleeding history. Diabetes, heart failure, female sex, and OAC therapy ranked relatively low in feature importance. Apart from minor differences in ranked feature importance, the risk predictors were similar in those people not on OAC therapy (Figure 5B).

For prediction of major bleeding, the most important risk features in ranked order were OAC therapy, age (continuous), eGFR grade, comorbidity number, hemoglobin level, antiplatelet drug use, heart failure, length of hospital stay, and dementia (Figure 5C). With the exclusion of patients on OAC therapy, the more important risk features for bleeding were similar, except that use of gastric protection agents was a more highly ranked risk feature (Figure 5D).

For prediction of all-cause death, the most important risk features in ranked order were age (continuous), dementia, comorbidity number, heart failure, hemoglobin level, eGFR grade, anaemia, and length of hospital stay (Figure 5E). Results were similar in the sub-cohort not on OAC therapy (Figure 5F).

**DISCUSSION**

In this study, we compared the performances of currently used risk stratification scores for predicting ischemic stroke and major bleeding risk with ML models. ML achieved a similar discriminant performance for predicting ischemic stroke. However, it has significantly improved discrimination for major bleeding risk in patients with AF. Moreover, all ML models achieved better calibration. Our sensitivity analysis, taking into account only patients discharged without oral anticoagulants, yielded results consistent with the main analysis. The ability to accurately identify the overlapping risk of ischaemic stroke and major bleeding in patients with AF has important potential to support clinicians in decision making as to those patients who would derive net clinical benefit from continuance or initiation of anticoagulation therapy while alerting to those who need greater attention to modifiable bleeding risk factors.

Our statistical measures of performance scores (AUC) are consistent with previously published studies.[28–30] The AUCs for prediction of ischemic stroke with the $CHA_2DS_2$-VASc scores were 0.67 (0.67-0.68) in a Swedish AF cohort study,[29] and 0.661 (0.633-0.690) in a nationwide cohort study in Denmark.[28] The AUCs for major bleeding risk prediction ranged from 0.57 to 0.61 in AF cohort studies.[30] Compared to $CHA_2DS_2$-VASc, HAS-BLED, ATRIA and ORBIT risk stratification schemas, the ML methods achieved better discrimination and calibration in predicting patients' one-year risk for ischemic stroke and major bleeding. Similar calibration was found between different multilabel ML models. Multilabel GBM showed the best discriminant power among different ML techniques, with estimated AUC improving by 7.4% for ischemic stroke and 7.5% for major bleeding. The NRI was not significant in predicting the risk of ischemic stroke, which means that ML models did not outperform $CHA_2DS_2$-VASc score in ischemic stroke risk stratification. This has cross-validated the importance of the risk factors already identified in the $CHA_2DS_2$-VASc schemas. Notably, the ML models, especially GBM, had moderately good discrimination for prediction of all-cause death and survival without stroke, with estimated AUC levels > 0.70 in the whole cohort as well as those not on anticoagulant therapy.

The ordering of class labels affects the predictive performance for both GBM and SVM classifier chains. We achieved the best performance by setting the major bleeding as the first class, followed by ischemic stroke and all-cause death, and the event-free survival as the last class. This ordering outperformed the ensemble of randomly ordered chains in predicting ischemic stroke and major bleeding in our data. The correlation between the patients' risks for major bleeding, ischemic stroke as well as all-cause death led to a superior predictive performance. Due to the effectiveness of anticoagulant therapy, patients who were admitted for major bleeding were less likely to have ischemic stroke events [8] with only 0.2% of patients in the total cohort admitted for both ischemic stroke and major bleeding during the follow-up period. However, both ischemic stroke and major bleeding could elevate mortality risk. By taking correlations between different outcomes into account, the multilabel ML models emerged as more robust risk predictor compared to the clinical risk scores that focuses on only one risk at a time.

In addition to pre-identified risk factors, the ML models can also automatically identify further potentially important risk factors without carrying out extensive studies. Risk factors identified by GBM overlapped with those identified by previously reported studies, which used standard statistical methods.[1,4,7] For example, it identified top-ranked features such as age and stroke history, which are also important risk factors in $CHA_2DS_2$-VASc and HAS-BLED scores. Previously studies have also confirmed that people with dementia and heart failure have substantially increased the risk of mortality.[31,32] This verifies that our ML model is reliable in identifying risk factors. GBM has identified other potential risk factors for ischemic stroke, such as the number of coded comorbidities, cardiology admission, and hemoglobin result. Other than the risk factors in the major bleeding risk stratification schemas, it also identified renal function (eGFR grade) and hemoglobin level as important predictors for major bleeding. Not surprisingly, OAC use was identified as an important feature for major bleeding, hence the drop of AUC when the ML algorithm was applied to the sub-cohort not on OAC therapy. Nevertheless, the ML model excluding OAC use was associated with a moderately large NRI (11.6%) when compared with the best clinical bleeding risk score.

Ischemic stroke and major bleeding risk prediction have become increasingly important in clinical decision-making. ML approaches may offer improved accuracy and more personalized risk assessment. This may also support the drive towards rationalized OAC therapy for patients with AF. The improved accuracy achieved in the current study should be further explored using ML techniques with other large linked clinical datasets. Further investigation of the feasibility and acceptability of ML in clinical decision-making on the prescription of anticoagulation therapy will be needed. As a result of the availability of large-scale data and improved computational capacity, ML techniques are expected to enhance the prediction of stroke and bleeding risk in patients with AF.

**Limitations**

The study has several limitations. First, we tested only the performance of the simple clinical risk scores currently used by clinicians to predict ischemic stroke and major bleeding risk. One study showed that ML techniques did not improve risk prediction when compared to stepwise logistic regression models using all potential predictors rather than just simple scores used in routine clinical practice such as $CHA_2DS_2$-VASc and HAS-BLED.[9] We only focused on whether the patient had an event or not during the follow-up period, and ignored the order of event, as we were unable to adjust patients' risk if they experienced any event during the follow-up period. This limitation can be addressed in the future using more complex multilabel modelling regarding time variance. Third, our cohort is the hospital population, which is older compared to the general AF population, which may introduce potential bias into our study. Moreover, due to the inherent limitations of the database, this study uses a modified HAS-BLED score. Notably, labile INR was excluded from the score, and this may alter the scores performance characteristics. We ignored the cause the death in our analysis as not all patients in the cohort have a cause of death code. Patients whose death were caused by ischemic stroke or major bleeding were classified as all-cause death. This may affect the accuracy of our mortality. Lastly, multiple potential features such as troponin and calcium with more than 50% missing values were excluded from our study. The improvement of our ML models could be more significant with this extra information available.

## CONCLUSIONS

Compared to established risk stratification schemes, this study has shown that multilabel ML techniques can predict ischemic stroke and major bleeding risk in patients with AF and had better discrimination power for prediction of major bleeding and all-cause mortality. GBM classifier chain achieved the best performance compared to SVM and multilabel MLNN. Furthermore, GBM were also able to identify important risk factors in addition to those used by traditional risk stratification tools.

**Contributorship Statement**

JL contributed with data processing, statistical and machine learning analysis, interpretation of results, and drafted & revised the manuscript. JH advised on the dataset and analytical plan and contributed to the interpretation of results and revision of the manuscript. FS and MB advised on machine learning and revised the manuscript. KM helped with the statistical analysis. BC and JS helped with the manuscript revision. RH, TB, and BMQ were involved in the acquisition of data, interpretation of data and revised the manuscript. FMS and GD conceived the study and provided clinical, epidemiological and statistical advice and interpretation, and revised the manuscript.

**Acknowledgments**

We thank the following institutions for providing the data used in this study. Staff at the WA Data Linkage Branch and data custodians of the WA Department of Health Inpatient Data Collections and Registrar General for access to and provision of the State linked data.

**Sources of funding**

This work was supported by a project grant from the Western Australian Department of Health State Health Research Advisory Council (Application 2011–154).

**Competing interests**

The authors declare no conflicts related to the contents published in this article.

**Data Availability Statement**

We will consider requests for data sharing on an individual basis, with the aim to share data whenever possible for appropriate research purposes. However, this research project uses data obtained from a third-party source under strict privacy and confidentiality agreements from the Sir Charles Gairdner and Osborne Park teaching hospitals in Perth, Western Australian Department of Health (State) and Australian Department of Health (Federal) databases, which are governed by their ethics committees and data custodians. The data were provided after approval was granted from their standard application processes for access to the linked datasets. Therefore, any requests to share these data with other researchers will be subject to formal approval from the third-party ethics committees and data custodian(s). Researchers interested in these data should contact the Client Services Team at the Data Linkage Branch of the Western Australian Department of Health (www.datalinkage-wa.org.au/contact-us).

**Figure Legends**

Figure 1. Proposed framework

Figure 2. Flow chart of the study population

Figure 3. Multilabel machine learning architecture

Figure 4. ROC Curves for GBM and CHADS-VASc and HAS-BLED scores

A, ROC Curves for GBM and CHADS-VASc and HAS-BLED score predicting ischemic stroke and major bleeding. B, ROC Curves for GBM and $CHA_2DS_2$-VASc and HAS-BLED scores predicting ischemic stroke and major bleeding for patients not on OAC.

ROC, receiver operating characteristics; GBM, gradient boosting machine; OAC, oral anticoagulants; $CHA_2DS_2$-VASc, congestive heart failure, hypertension, age ≥ 75, diabetes mellitus, transient ischemic attack or stroke or systemic arterial thromboembolism, vascular disease, age 65-74, sex; HAS-BLED, hypertension, age, stroke, bleeding history, liver and kidney dysfunction, elevated INR, drugs and alcohol

Figure 5. Top 25 important variables for predicting different outcomes.

A, ischemic stroke; B, Ischemic stroke, no OAC; C, Major bleeding; D, Major bleeding, no OAC; E, All-cause mortality and F, All-cause mortality, no OAC; ranked by gradient boosting machine in the overall AF cohort and sub-cohort not on OAC therapy.

AF, atrial fibrillation; OAC, oral anticoagulant; eGFR, estimated Glomerular Filtration Rate; ACE, angiotensin-converting enzyme; ARB, angiotensin receptor blocker; AF, atrial fibrillation. DVT, deep vein thrombosis; PE, pulmonary embolus.

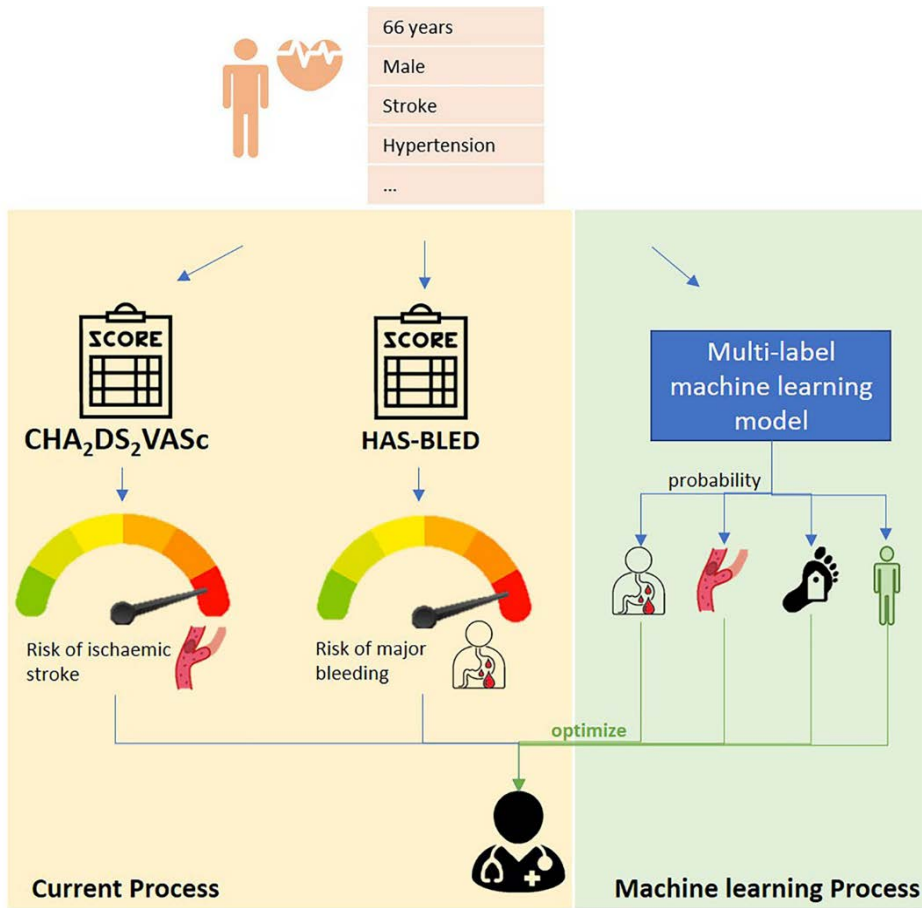

Figure 1

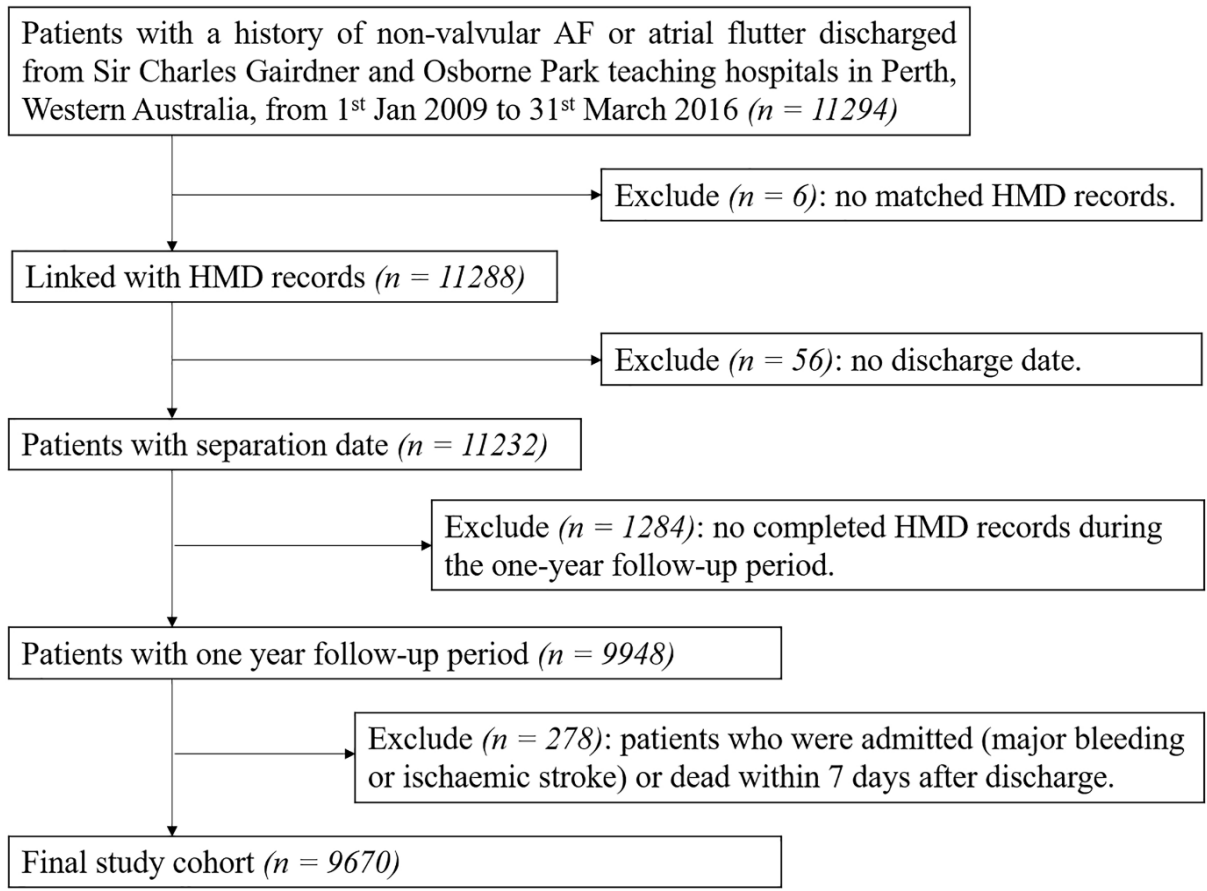

Figure 2

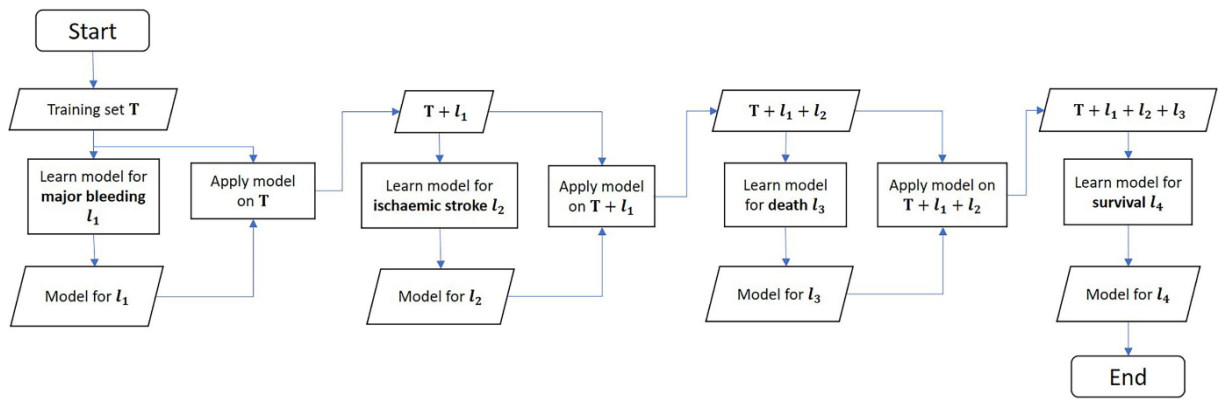

Figure 3

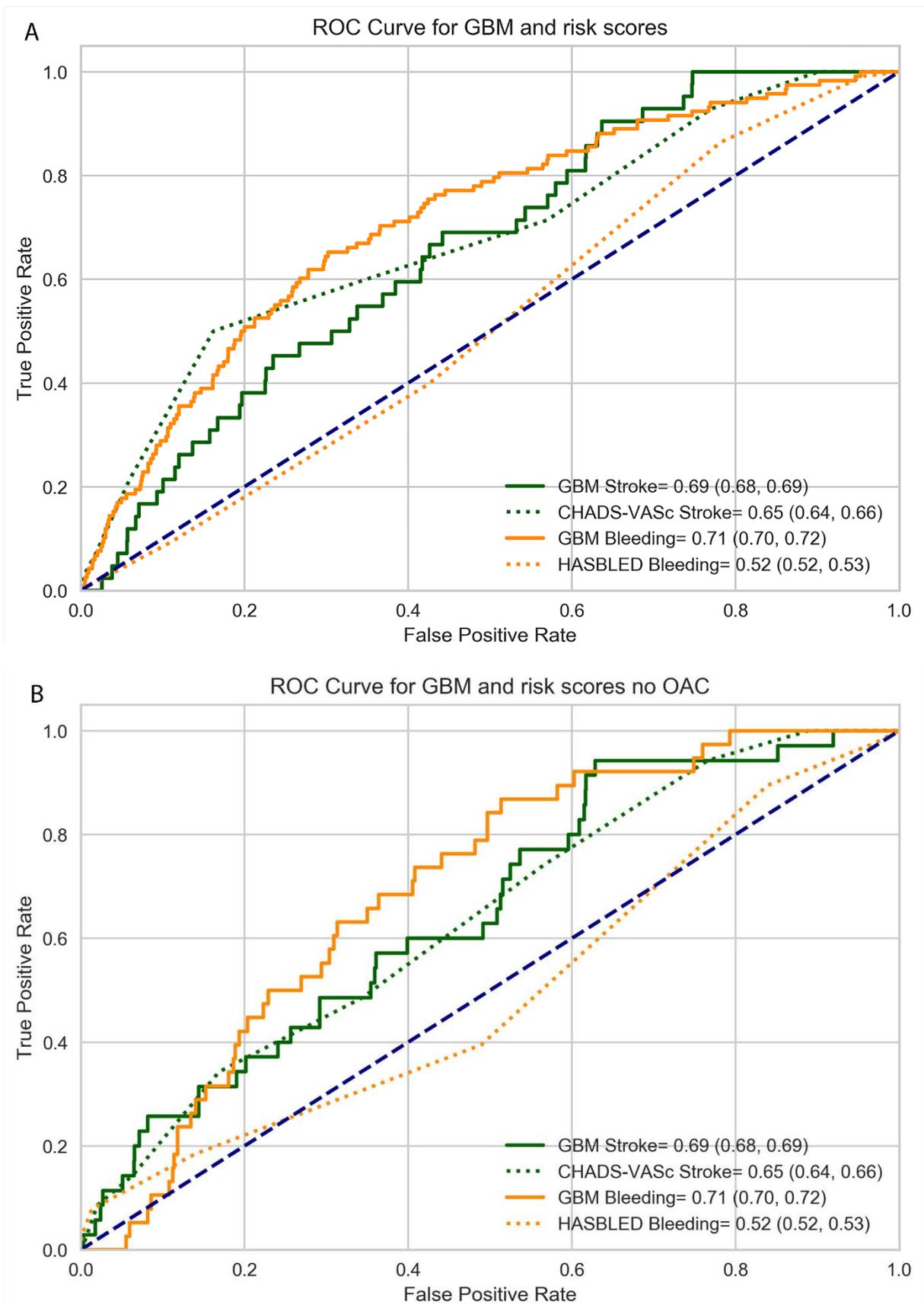

Figure 4

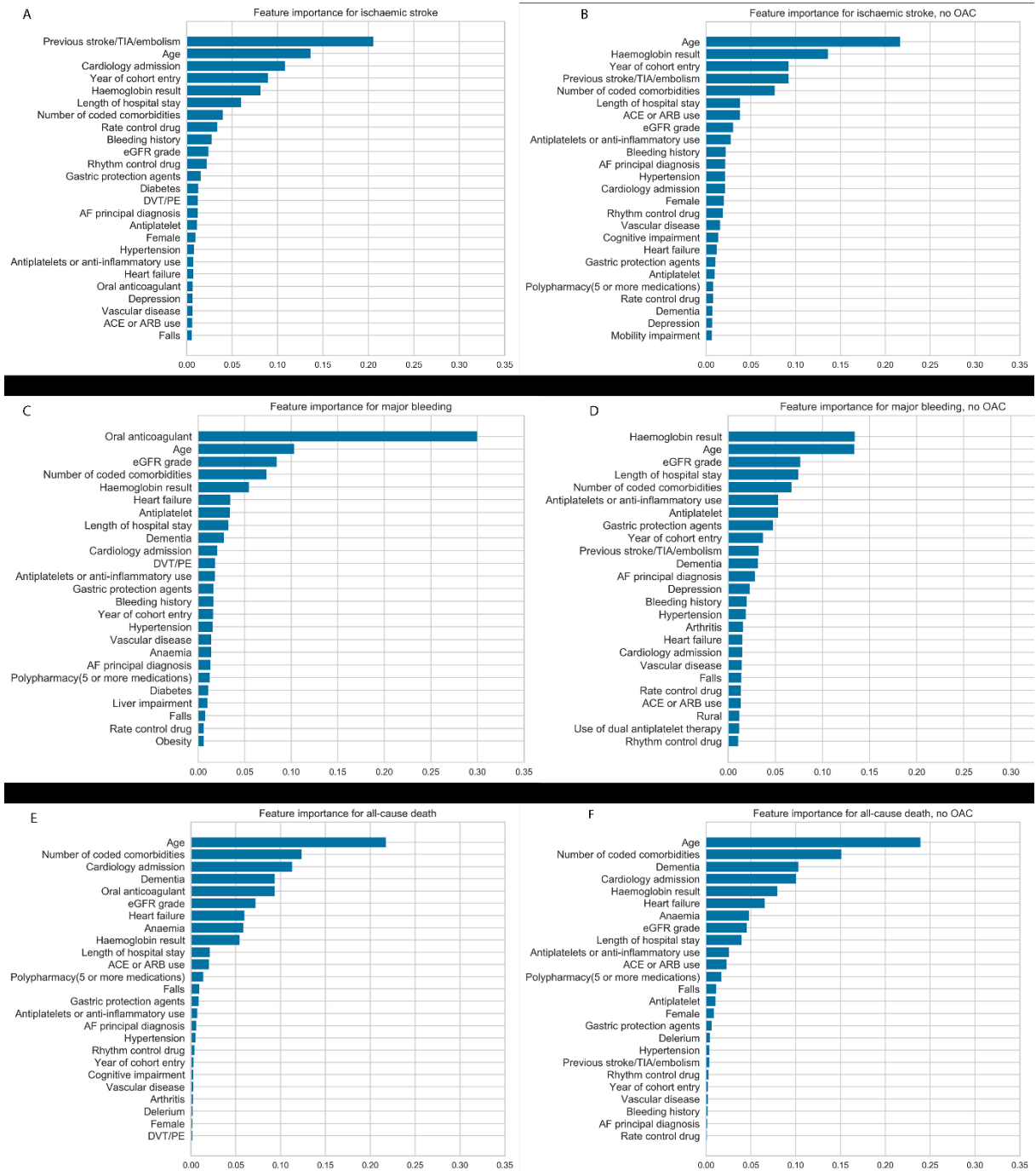

Figure 5